\documentclass[10pt,conference]{IEEEtran}
\IEEEoverridecommandlockouts
\usepackage{cite}
\usepackage{amsmath,amssymb,amsfonts}
\usepackage{algorithmic}
\usepackage{graphicx}
\usepackage{textcomp}
\usepackage{xcolor}
\usepackage{tcolorbox}
\usepackage{multirow}
\usepackage{booktabs}
\usepackage{pifont}
\usepackage{threeparttable}
\def\BibTeX{{\rm B\kern-.05em{\sc i\kern-.025em b}\kern-.08em
    T\kern-.1667em\lower.7ex\hbox{E}\kern-.125emX}}
\begin{document}

\title{LLaMA-Reviewer: Advancing Code Review Automation with Large Language Models through Parameter-Efficient Fine-Tuning
\thanks{$^{\ast}$Corresponding author.}}

\author{
  \IEEEauthorblockN{
        Junyi Lu\IEEEauthorrefmark{2}\IEEEauthorrefmark{3}, 
        Lei Yu\IEEEauthorrefmark{2}\IEEEauthorrefmark{3},
        Xiaojia Li\IEEEauthorrefmark{4},
        Li Yang\IEEEauthorrefmark{1}\IEEEauthorrefmark{2},
        Chun Zuo\IEEEauthorrefmark{5}
    }
  \IEEEauthorblockA{
    \IEEEauthorrefmark{2}Institute of Software, Chinese Academy of Sciences, Beijing, China \\
    \IEEEauthorrefmark{3}University of Chinese Academy of Sciences, Beijing, China \\
    \IEEEauthorrefmark{4}School of Software, Tsinghua University, Beijing, China
    \IEEEauthorrefmark{5}Sinosoft Company Limited, Beijing, China \\
    \{lujunyi21, yulei21\}@mails.ucas.ac.cn, lixj21@mails.tsinghua.edu.cn, \\
    yangli2017@iscas.ac.cn, zuochun@sinosoft.com.cn
    }
} 


\maketitle

\begin{abstract}
The automation of code review activities, a long-standing pursuit in software engineering, has been primarily addressed by numerous domain-specific pre-trained models. Despite their success, these models frequently demand extensive resources for pre-training from scratch. In contrast, Large Language Models (LLMs) provide an intriguing alternative, given their remarkable capabilities when supplemented with domain-specific knowledge. However, their potential for automating code review tasks remains largely unexplored.

In response to this research gap, we present LLaMA-Reviewer, an innovative framework that leverages the capabilities of LLaMA, a popular LLM, in the realm of code review. Mindful of resource constraints, this framework employs parameter-efficient fine-tuning (PEFT) methods, delivering high performance while using less than 1\% of trainable parameters.

An extensive evaluation of LLaMA-Reviewer is conducted on two diverse, publicly available datasets. Notably, even with the smallest LLaMA base model consisting of 6.7B parameters and a limited number of tuning epochs, LLaMA-Reviewer equals the performance of existing code-review-focused models.

The ablation experiments provide insights into the influence of various fine-tuning process components, including input representation, instruction tuning, and different PEFT methods. To foster continuous progress in this field, the code and all PEFT-weight plugins have been made open-source.
\end{abstract}


\section{Introduction}

Since its formalization by Fagan in 1976 \cite{fagan2002design}, code review has been a cornerstone of software engineering, instrumental in defect identification, quality improvement, and knowledge sharing \cite{spadini2020primers}. However, this primarily manual process imposes a significant workload on developers. Even with modern code review (MCR) practices, which are more streamlined than traditional ones, the effort required is still substantial \cite{rigby2014peer, sadowski2018modern, shan2022using}.

To alleviate this burden, a surge of research has focused on automating the code review process. This includes tasks such as recommending reviewers \cite{sulun2019suggesting, asthana2019whodo, chueshev2020expanding, rebai2020multi, mirsaeedi2020mitigating, sulun2021rstrace+, gauthier2021historical, kong2022recommending, rong2022modeling, pandya2022corms}, evaluating code quality \cite{shi2019automatic, hellendoorn2021towards, hijazi2021ireview, wang2021can, gauthier2021historical, li2022automating, hijazi2022quality}, refining problematic code \cite{tufano2021towards, tufano2022using, fu2022vulrepair, li2022automating, thongtanunam2022autotransform}, and generating potential review comments \cite{balachandran2013reducing, gupta2018intelligent, siow2020core, hong2022should, tufano2022using, li2022automating, li2022auger, hong2022commentfinder}. Recent advancements in natural language processing (NLP) have further enabled the use of pre-trained language models (PLMs) for these tasks \cite{tufano2022using, li2022automating}. However, such domain-specific models often require substantial resources for pre-training from scratch.

In contrast, unified large language models (LLMs) demonstrate remarkable performance when scaled to a certain parameter size \cite{gauthier2021historical, kong2022recommending}. They can effectively handle specific tasks without the need for domain-specific pre-training, presenting a promising avenue for code review automation.

In this study, we present LLaMA-Reviewer, a novel framework that leverages LLaMA, a mainstream LLM, for automating code review. We incorporate Parameter-Efficient Fine-Tuning (PEFT) methods to address the computational challenge of LLM fine-tuning. Our approach builds upon the pipeline proposed by Li et al. \cite{li2022automating}, which comprises 1) review necessity prediction, 2) review comment generation, and 3) code refinement tasks.

We extensively evaluate LLaMA-Reviewer on two public datasets for each sub-task and investigate the impacts of the input representation, instruction tuning, and different PEFT methods. The primary contributions of this work include:

\begin{itemize}
\item Introducing the application of LLMs to code review automation tasks, offering an offline and privacy-conscious alternative to closed-source solutions like OpenAI APIs.
\item Proposing a ``unified model + PEFT" paradigm to reduce computational demands during code review tasks, with the plug-in model being part of it to optimize storage space requirements, first in software engineering domain.
\item Conducting a comprehensive evaluation of two PEFT methods and ablation studies on fine-tuning components.
\item Open-sourcing our code, models, and results\cite{anonymous_2023_7991113}.
\end{itemize}

Here is the paper's structure: Section II gives the necessary background; Section III details our proposed approach; Section IV describes the experiment design; Section V discusses evaluation results; Section VI reviews related work; Section VII identifies potential validity threats; Section VIII concludes our findings and proposes future research directions.

\section{Background}
This section provides a succinct introduction to three key concepts underlying our research: the automated code review pipeline, large language models (LLMs), and parameter-efficient fine-tuning (PEFT) methods.

\subsection{Automation in Code Review}

\begin{figure}
\centering
\includegraphics[width=\linewidth]{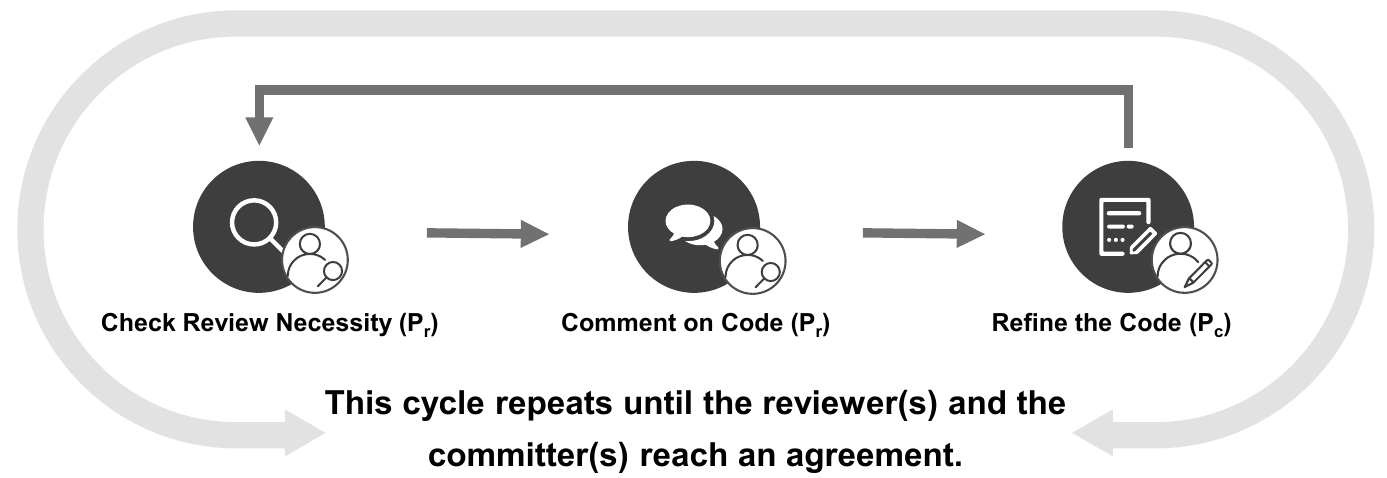}
\caption{The cycle of the code review process.}
\label{fig: code review process}
\end{figure}

Modern Code Review (MCR), a technique extensively adopted by both large enterprises and open-source projects, has a relatively consistent core cycle despite varied implementations. This cycle, from the creation of a pull request to its final merge into the main branch or rejection, involves two primary participants: the committers ($P_c$) and reviewers ($P_r$). The cycle encompasses three key steps (as shown in Figure \ref{fig: code review process}): review necessity prediction ($P_r$), commenting on code ($P_r$), and code refinement ($P_c$). The objective of automating the code review process is to alleviate the workload for both parties.

The three steps translate into three automation tasks: \textbf{1) Review Necessity Prediction}, predicting whether a code diff requires review comments; \textbf{2) Review Comment Generation}, automatically generating review comments for a given code diff; and \textbf{3) Code Refinement}, refining the code based on prior snippets and review comments. Our study focuses on these tasks, aiming to fully automate the code review process.

\subsection{Large Language Models}

The evolution of language modeling (LM) has seen four significant stages: statistical language models (SLMs), neural language models (NLMs), pre-trained language models (PLMs), and the latest development, large language models (LLMs) \cite{zhao2023survey}. PLMs, explicitly pre-trained for certain tasks, have been notably successful in various downstream software engineering tasks. This is demonstrated by models like CodeT5 \cite{wang2021codet5} and PLBART \cite{ahmad2021unified}. However, the potential of LLMs in these contexts has yet to be fully explored.

The primary distinction between PLMs and LLMs is their scale of parameters and data size. LLMs are models with $\sim$10B parameters or more, pre-trained with extensive data\cite{zhao2023survey}. Current research suggests that scaling these dimensions improves model performance and gives rise to emergent capabilities \cite{wei2022emergent}. Notably, LLMs can achieve performance on par with PLMs without the necessity for task-specific pre-training, thus alleviating the resource-heavy demands of pre-training.

To be more specific, the currently trending LLMs can be categorized into unified LLMs and code LLMs. The formers are predominantly pre-trained on natural language corpora, enriched with a smaller portion of code, and have been validated as effective in various tasks\cite{touvron2023llama, openai2023gpt4}. The latters are primarily pre-trained on a code-based corpus and they have achieved impressive results in code generation~\cite{fried2022incoder, li2023starcoder, nijkamp2022codegen, wang2023codet5+}. 

In this research, we utilize LLaMA, an open-source unified LLM developed by Meta. This choice stems from four perspectives: 1)Top-performing models (GPT-3.5/GPT-4) for code tasks are unified models, not code LLMs; 2)The increasing trend towards unified models, exemplified by OpenAI's transition from Codex to GPT-3.5 for API usage; 3) The increasing trend towards unified models, exemplified by OpenAI's transition from Codex to GPT-3.5 for API usage; 4) Code LLMs primarily excel in code generation tasks, whereas code review tasks pose different challenges.

\subsection{Parameter-Efficient Fine-Tuning}

Despite their effectiveness, the high computational resources required for fine-tuning large language models (LLMs) present a significant challenge. Numerous strategies have been developed to increase the efficiency of the fine-tuning process and lower training costs. These include adapter tuning \cite{houlsby2019parameter,he2021towards}, prefix tuning \cite{li2021prefix}, prompt tuning \cite{lester2021power,liu2021gpt}, and low-rank adaptation (LoRA) \cite{hu2021lora}. These methods freeze the base model's parameters while training a few additional parameters, achieving performance comparable to full-parameter tuning.

In this study, we use two PEFT methods—zero-init attention prefix-tuning\cite{zhang2023llama} and LoRA tuning\cite{hu2021lora}—to fine-tune LLaMA. These methods do not introduce extra latency to the model and have demonstrated effectiveness in natural language tasks. The PEFT methods' specifics are further elaborated in the subsequent section on our proposed approaches.

\section{LLaMA-Reviewer: Proposed Approach}

\subsection{Overview}
\begin{figure}
\centering
\includegraphics[width=\linewidth]{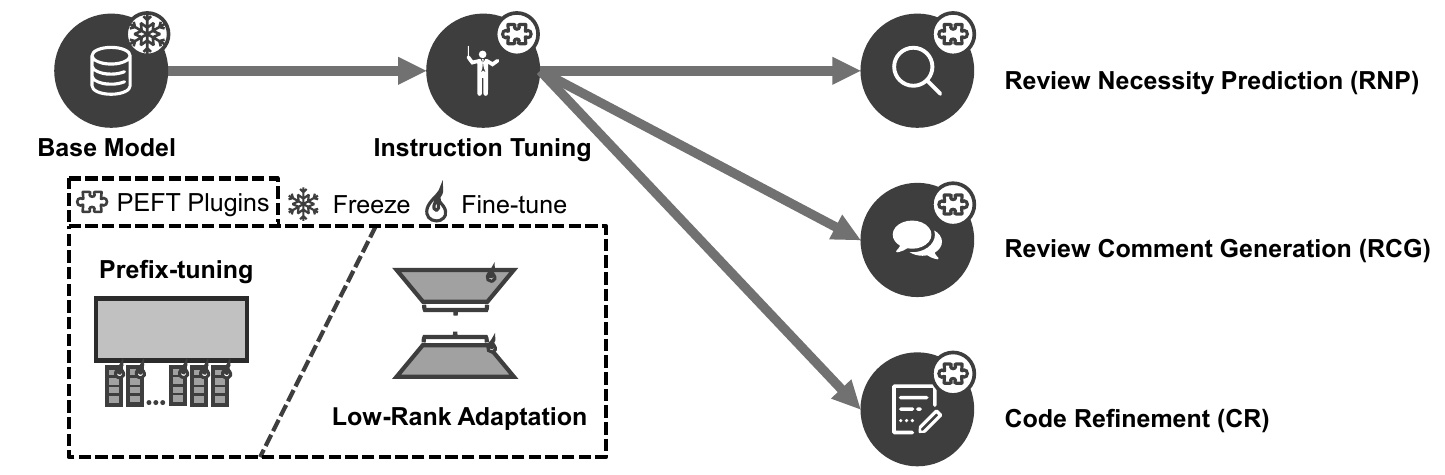}
\caption{The overview of LLaMA-Reviewer.}
\label{fig: overview}
\end{figure}

Our framework, illustrated in Figure \ref{fig: overview}, employs a dual-stage fine-tuning process. We initiate with instruction-following tuning on LLaMA using code-centric domain data, enhancing the model's proficiency in comprehending code review tasks and adhering to task instructions. We then conduct supervised fine-tuning for each sub-task within the code review process using the enhanced LLaMA. To balance parameter efficiency and model performance, we incorporate two key techniques---zero-init attention prefix-tuning and low-rank adaptation (LoRA) tuning---since they have achieved wide acceptance and promising results, especially with LLaMA \cite{Tloen, zhang2023llama}. These methods, founded on the plugin encapsulation strategy of PEFT, furnish us with lightweight task-specific plugins. These plugins, independent of the base model's weights, can be seamlessly integrated during inference.

We evaluate the effectiveness of our approach using two distinct datasets. For clarity, we will refer to the dataset in CodeReviewer\cite{li2022automating} as the ``CRer dataset", and the dataset from Tufano et al.\cite{tufano2022using} as the ``Tuf. dataset". Further details on the evaluation process can be found in the subsequent sections.

\subsection{Instruction Tuning on LLaMA}
Research indicates that when LLMs are fine-tuned on a diverse range of multi-task datasets using natural language descriptions, they demonstrate enhanced performance on unseen tasks \cite{wei2021finetuned,ouyang2022training}. Instruction-following tuning aids the model in better interpreting user intentions and following instructions. To initially adapt the pre-trained model for code review tasks, we employ instruction tuning on a code-centric domain.

\begin{figure}
\centering
\includegraphics[width=\linewidth]{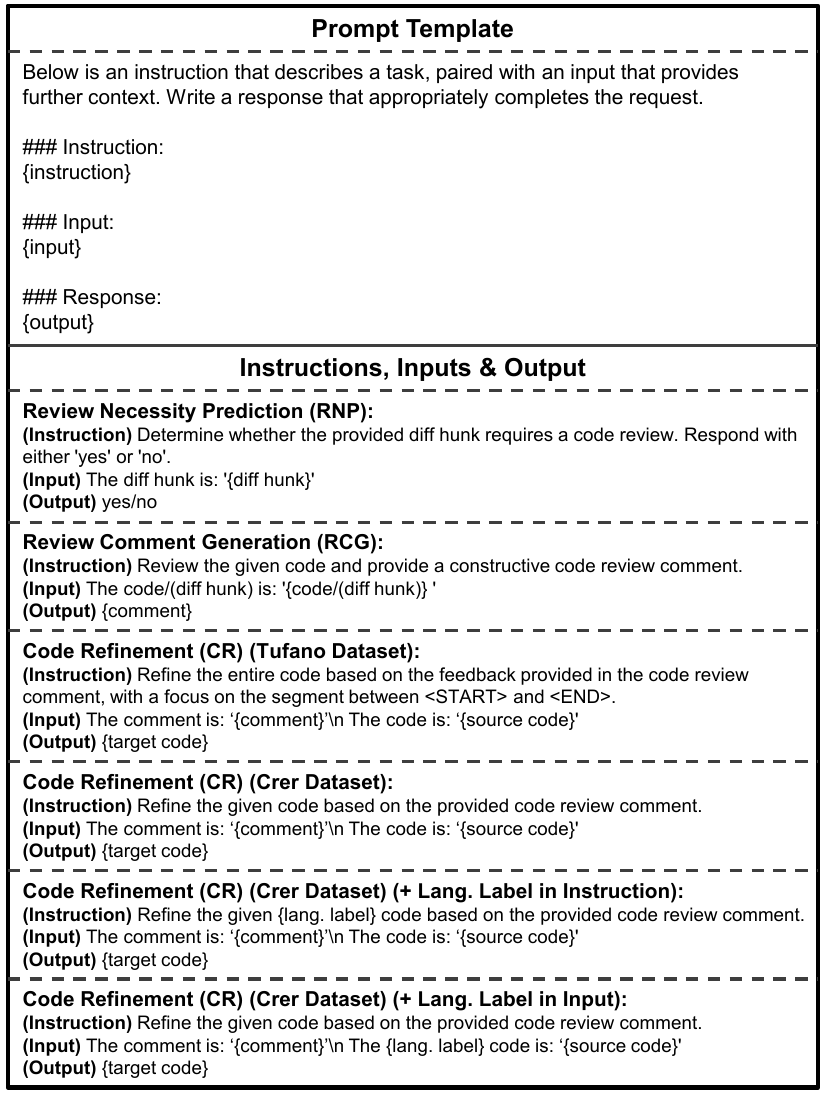}
\caption{The prompt template and instruction, input and output formats.}
\label{fig: prompt}
\end{figure}

We leverage the primary process and template from Stanford Alpaca \cite{alpaca}, modifying the full-parameter fine-tuning to use PEFT methods as explained in Section III-C and Section III-D. Given our code review task's deep relevance to coding, we substitute the original data with its code-domain equivalent, Code Alpaca \cite{codealpaca}. Combining data from Alpaca and Code Alpaca was considered but did not lead to performance improvements, as further discussed in the ablation experiments section. The data structure adheres to the \{instruction, input (optional), output\} format, following the framework from \cite{wang2022self}. The same prompt template is used for subsequent sub-tasks to maximize the use of the first stage fine-tuned model. Figure \ref{fig: prompt} illustrates the prompt template and sub-task instruction, input, and output formats.

\subsection{Zero-init Attention Prefix-tuning}

Zero-init attention prefix-tuning, an offshoot of prefix-tuning, keeps the base model's weights intact while integrating an extra $K$ prefix tokens into the topmost $L$ layers of LLaMA's transformer. These flexible prompts are concatenated with the original tokens, allowing the computation of multi-head attention using the newly introduced keys and values. During the fine-tuning, only these adaptable prompts are trained.

This method deviates from conventional prefix-tuning by introducing a learnable gating factor during attention calculation. This factor regulates the relevance of the inserted prompt tokens. To facilitate the tuning process, this gating factor is initially zeroed, leading to the `zero-init' in the name of the method. This factor specifically controls the attention shared between the prompt tokens and the original tokens.

Figure \ref{fig: prefix-tuning} illustrates the details. We take the $l$ layer, one of the topmost $L$ layers as an example. Here, $P_{l} \in \mathbb{R}^{K \times C}$ represents the $K$ prompt tokens and $T_{l} \in \mathbb{R}^{M \times C}$ signifies the $M$ original tokens in the $l$ layer. The feature dimension is symbolized by $C$. When the $(M+1)$-th token $t_{M+1}$ is computed via the attention mechanism, the queries, keys, and values are obtained through several linear layers as follows:

\begin{figure}
\centering
\includegraphics[width=\linewidth]{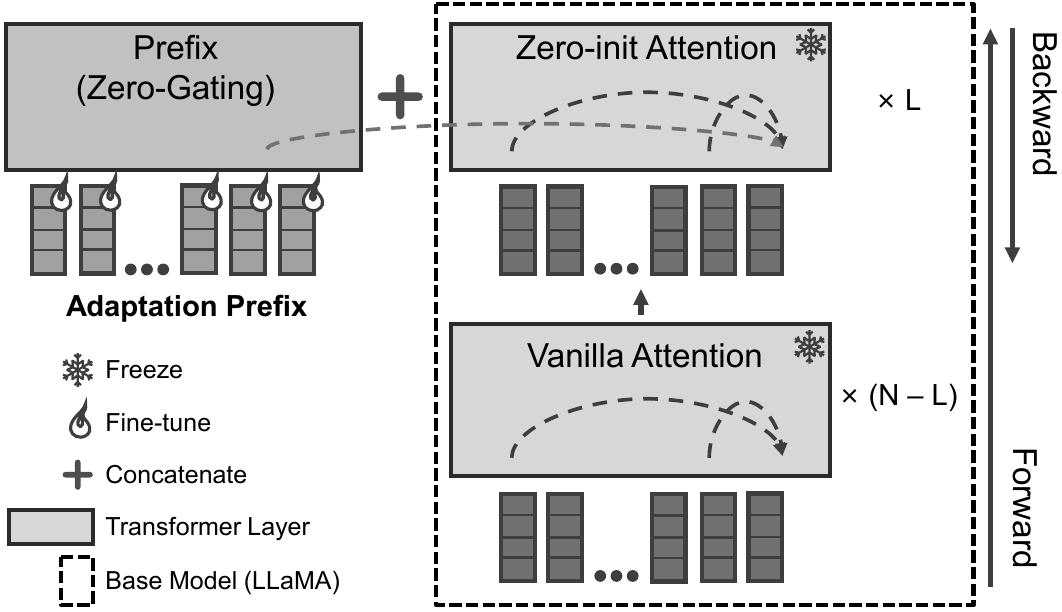}
\caption{The details of prefix-tuning on LLaMA.}
\label{fig: prefix-tuning}
\end{figure}

\begin{equation}
\begin{aligned}
Q_{l} & =\operatorname{Linear}_{\mathrm{q}}\left(t_{M+1}\right) \\
K_{l} & =\operatorname{Linear}_{\mathrm{k}}\left(\left[P_{l} ; T_{l} ; t_{M+1}\right]\right) \\
V_{l} & =\operatorname{Linear}_{\mathrm{v}}\left(\left[P_{l} ; T_{l} ; t_{M+1}\right]\right)
\end{aligned}   
\end{equation}

Next, the attention scores are derived:

\begin{equation}
\begin{aligned}
S_{l} & =\left[S_{l}^{K} ; S_{l}^{M+1}\right]^{T} \\
S_{l}^{K} & =Q_{l} \left(K_{l}^{K}\right)^{T} / \sqrt{C} \in \mathbb{R}^{1 \times K} \\
S_{l}^{M+1} & =Q_{l} \left(K_{l}^{M+1}\right)^{T} / \sqrt{C} \in \mathbb{R}^{1 \times (M+1)}
\end{aligned}
\end{equation}

The gating factor is integrated post-softmax application:

\begin{equation}
   S_{l}^{g}=\left[\operatorname{Softmax}\left(S_{l}^{K}\right) \cdot g_{l} ; \operatorname{Softmax}\left(S_{l}^{M+1}\right)\right]^{T} 
\end{equation}

Ultimately, the output of the $t_{M+1}$ token is produced via a linear projection layer, and it is controlled by prefixes:

\begin{equation}
    t_{M+1}^{o}=\operatorname{Linear}_{o}\left(S_{l}^{g} V_{l}\right) \in \mathbb{R}^{1 \times C} .
\end{equation}

\subsection{Low-Rank Adaptation}
Low-Rank Adaptation (LoRA) provides a different perspective on Parameter-Efficient Fine-Tuning (PEFT). Unlike full-parameter tuning methods that require all weights to be updated during the fine-tuning phase, LoRA retains the weights of the original model and integrates trainable low-rank matrices to the transformer layers to simulate the weight adjustments. This approximation draws on the principle that the adaptation process inherently has a low ``intrinsic rank."

Figure \ref{fig: lora} illustrates the core component of LoRA. Suppose $W_{0} \in \mathbb{R}^{d \times k}$ represents the pre-trained matrix. The approximation of the weight adjustment from $W_{0}$ to $W_{0} + \Delta W$ using LoRA can be expressed as:

\begin{figure}
\centering
\includegraphics[width=0.8\linewidth]{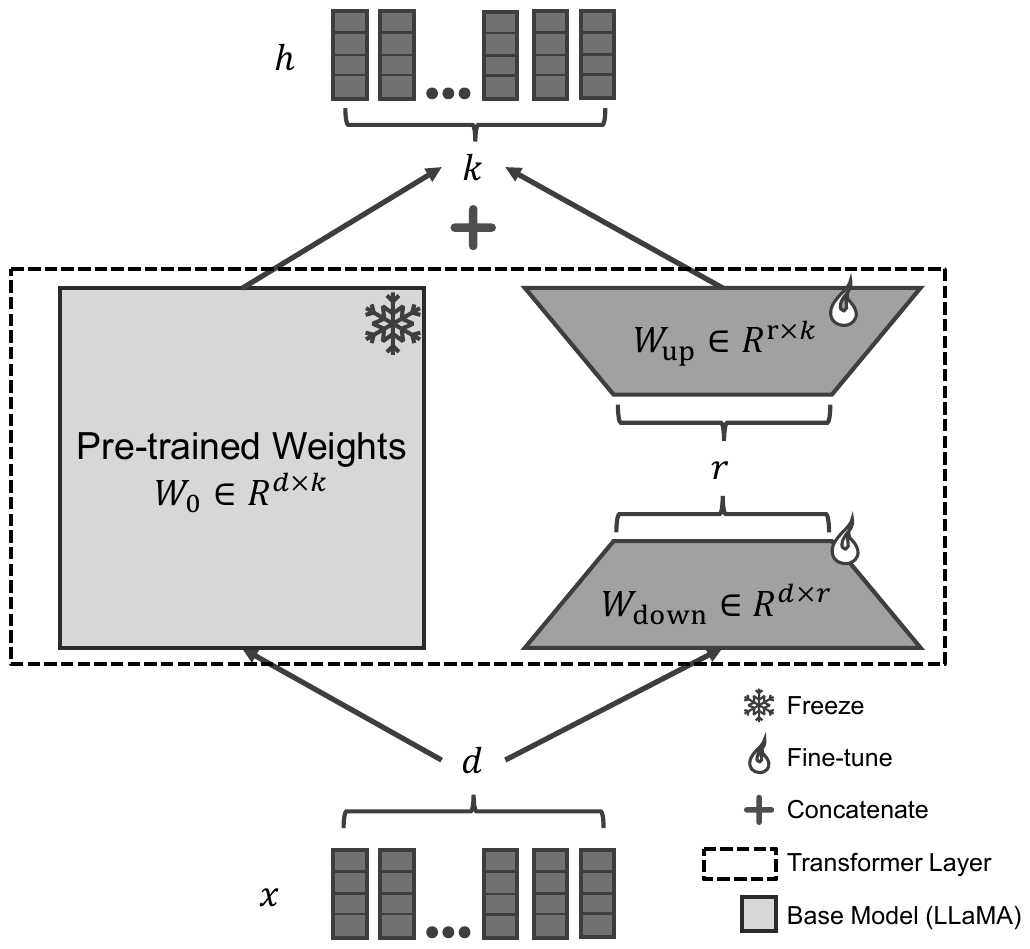}
\caption{The core component of Low-Rank Adaptation (LoRA).}
\label{fig: lora}
\end{figure}

\begin{equation}
    W_{0} + \Delta W = W_{0} + W_{down}W_{up}
\end{equation}

Here, $W_{down} \in \mathbb{R}^{d \times r}$ and $W_{up} \in \mathbb{R}^{r \times k}$, with $r \ll \min (d, k)$ being the rank. During the fine-tuning process, $W_{0}$ remains unchanged, whereas $W_{down}$ and $W_{up}$ become the trainable parameters. Given an input $x$ and its associated original output $h$, the adjusted output $\bar{h}$ is computed as:

\begin{equation}
    \begin{aligned}
        \bar{h} &= W_{0} x + \Delta W x = h + W_{down}W_{up} x
    \end{aligned}
\end{equation}

\subsection{Code Review Automation Tasks}
LLaMA-Reviewer is specifically designed to automate three core tasks integral to the code review process, namely review necessity prediction, code review comment generation, and code refinement. These tasks sequentially correspond to stages in a typical code review process. Inputs and outputs for these tasks are categorized into three formats:
\begin{itemize}
\item Programming Language (PL) for code snippets,
\item Natural Language (NL) for code comments, and
\item Binary Labels (L) for decisions on the requirement for further review. This simplifies the decision process to a ``yes" (review needed) or ``no" (review not needed).
\end{itemize}

Table \ref{tab: tasks} illustrates each task along with its input and output formats, and corresponding dataset references (Crer. for the CRer dataset, and Tuf. for the Tufano dataset). The prompt construction approach is visualized in Figure \ref{fig: prompt}.

\begin{table}
\centering
\caption{Summary of the Code Review Automation Tasks}
\begin{tabular}{@{}cccc@{}}
\toprule
Task & Input & Output & Datasets \\ \midrule
Review Necessity Prediction & PL & L (yes/no) & Crer. \\
Code Review Comment Generation & PL & NL & CRer., Tuf. \\
Code Refinement & PL, NL & PL & CRer., Tuf. \\ \bottomrule
\end{tabular}
\label{tab: tasks}
\end{table}

\subsubsection{Review Necessity Prediction}
This task involves checking if diff hunks need reviews. Proposed by Li et al.\cite{li2022automating}, a diff hunk represents a concise piece of code that shows differences between old and new code snippets. Although including additional method context could be beneficial, the lines in the original diff hunk are typically long enough to present a challenge when managed as input.

\subsubsection{Code Review Comment Generation}
This task generates pertinent comments for a given code snippet. Two perspectives are considered: a line-level perspective that focuses on the content of individual lines of code (using the CRer dataset), and a method-level perspective that provides a more holistic view of the code's context (using the Tufano dataset).

\subsubsection{Code Refinement}
Code refinement entails making minor adjustments or reordering the existing code to enhance its quality. Given the nature of these minor modifications, the input and output codes often bear strong similarities. Inputs are formatted according to the Tufano and CRer datasets. Typically, We omitted language type information as the baseline models did so and it is only available in one task of one published dataset, to ensure a fair comparison. The implications of such information are further explored in our evaluation section.
\section{Experimental Design}
This section details our experimental design, outlining the fundamental research questions driving our investigation, the datasets employed, the evaluation metrics used, and a thorough summary of the baseline models and our implementation specifics.

\subsection{Research Questions}
To evaluate the effectiveness of our proposed framework, we posit the following research queries:

\noindent\textbf{(RQ1) How effective is a large language model in automating code review tasks, compared to state-of-the-art methods?}

\noindent\textbf{Motivation.} The rapid advancements in AI-Generated Content (AIGC) and the known correlation between model capabilities and their size have led to the widespread use of fine-tuned, pre-trained large language models (LLMs). Although programming language data plays a significant role in augmenting a model's capabilities, the application of such data to code-related tasks—particularly those demanding proficiency in both natural language (NL) and programming language (PL), such as automated code review—remains largely unexplored. In this study, we employ the smallest variant of LLaMA as our base large language model to assess its effectiveness in automating code review tasks compared to state-of-the-art methods, notably task-specific pre-trained models like CodeReviewer\cite{li2022automating}. The model's performance is scrutinized across each task to identify its strengths and areas that warrant enhancement. Specifically, we pose the following questions:

\begin{itemize}
\item (RQ1.1) How effective is a large language model at \textbf{checking review necessity} (classification)?
\item (RQ1.2) How proficient is a large language model at \textbf{generating code review comments} (NL generation)?
\item (RQ1.3) How capable is a large language model at \textbf{refining code} based on comments (PL generation)?
\end{itemize}

\noindent\textbf{(RQ2) How does the representation of input data impact the performance of large language models?}

\noindent\textbf{Motivation.}\indent Given the fixed pre-training data format, there may be discrepancies between this format and that required for specific tasks. To evaluate the capabilities of the large language model, we delve into two critical factors:
\begin{enumerate}
\item Code Formatting. We evaluate the model's performance on raw code input and input with modified formatting (including eradicating consecutive spaces) to determine which can be handled more effectively.
\item Programming Language Labels. With various programming languages, we examine the impact of specifying the language type in the input and the importance of the label's location within the prompt template.
\end{enumerate}
In light of these considerations, we pose two sub-questions:
\begin{itemize}
\item (RQ2.1) How does code formatting influence the model's performance?
\item (RQ2.2) How do the inclusion and placement of programming language labels affect the model's performance?
\end{itemize}

\noindent\textbf{(RQ3) How does instruction tuning influence the performance of subsequent sub-tasks?}

\noindent\textbf{Motivation.}\indent Instruction tuning, the inclusion of code-related instructions in the initial stage, aims to infuse domain knowledge and help the model understand sub-tasks better. The effectiveness of this approach and its compatibility with various parameter-efficient fine-tuning (PEFT) methods remains a rich area of exploration. Additionally, given the dual nature of code review tasks involving natural language (NL) and programming language (PL), a comparison between using exclusively PL-related instructions and a mix of NL and PL-related instructions is warranted. Consequently, we ask:

\begin{itemize}
\item (RQ3.1) What is the impact of the initial instruction tuning stage on zero-init attention prefix-tuning?
\item (RQ3.2) How does the initial instruction tuning stage influence low-rank adaptation (LoRA)?
\item (RQ3.3) Which approach yields superior results in code review tasks: using a mix of NL and PL-related instructions or relying solely on PL-related instructions?
\end{itemize}

\noindent\textbf{(RQ4) What implications arise from different parameter-efficient fine-tuning (PEFT) methods?}

\noindent\textbf{Motivation.}\indent Two PEFT methods are employed to alleviate computational resource demands during fine-tuning. However, achieving an optimal balance between efficiency and effectiveness is challenging. Accordingly, we intend to analyze the results obtained from these two methods. Furthermore, in the context of low-rank adaptation (LoRA), the rank $r$ determines the number of trainable parameters; thus, we perform an ablation study to investigate the impact of rank $r$. Lastly, we compare the two methods concerning parameter count and storage space requirements with prior methods. To this end, we explore the following sub-questions:

\begin{itemize}
\item (RQ4.1) Which PEFT method performs better, and why?
\item (RQ4.2) Within LoRA, how does the rank $r$ influence trainable parameters and the overall performance?
\item (RQ4.3) How do the two PEFT methods compare to previous approaches in terms of parameter efficiency and storage space requirements?
\end{itemize}

\subsection{Datasets}
We utilize two prominent code review datasets: the dataset from CodeReviewer of Li et al.\cite{li2022automating} (hereafter the CRer dataset), and the dataset from Tufano et al. \cite{tufano2022using} (hereafter the Tufano dataset). The rationale for our choice includes:

\begin{itemize}
\item Unlike other datasets in the literature \cite{li2022auger, pornprasitd} that cover specific sub-tasks, the chosen datasets encompass the entire code review process.
\item Both the CRer and Tufano datasets are derived from a diverse range of repositories, providing broad coverage. This contrasts with other datasets \cite{li2022auger, pornprasitd} that draw from a limited repository pool, potentially leading to bias due to their restricted scope.
\item The Tufano dataset is an enhanced version compared to those used in earlier studies \cite{thongtanunam2022autotransform, tufano2021towards, zhang2022coditt5}, making it preferable for its recency and comprehensiveness.
\item Both datasets have been seminal in the field, each offering unique features that contribute to our study.
\end{itemize}

The CRer dataset, a multi-language corpus, is drawn from GitHub repositories and adheres to a diff-aware, line-grained format. It preserves inline comments and docstrings within code snippets and retains consecutive spaces. This dataset is divided into three sub-datasets, each dedicated to a specific aspect of code review: review necessity prediction, code review comment generation, and code refinement.

The Tufano dataset, in contrast, is language-specific (Java) and aggregates data from both GitHub and Gerrit. It uses a function-grained format, removes comments and consecutive spaces, and does not reflect differences between the associated commit and the base branch. For code refinement tasks, it denotes areas of focus within the comments using ``$\langle$START$\rangle$" and ``$\langle$END$\rangle$" markers. We utilize two subsets of this dataset for code review comment generation and code refinement.

Table \ref{dataset-statistics} offers a detailed statistical summary of datasets.

\begin{table*}
\caption{Statistical Overview of the Tufano Dataset and the CRer Dataset.}
\label{dataset-statistics}
\footnotesize
\begin{tabular}{@{}l|rrr|rrr|rrr|cccc@{}}
\toprule
\multirow{2}{*}{Dataset} & \multicolumn{3}{c|}{Review Necessity Prediction} & \multicolumn{3}{c|}{Review Comment   Generation} & \multicolumn{3}{c|}{Code Refinement} & \multirow{2}{*}{Lang \#} & \multirow{2}{*}{Gran.} & \multirow{2}{*}{\begin{tabular}[c]{@{}c@{}}Indent. \&\\      Consec. Spaces\end{tabular}} & \multirow{2}{*}{\begin{tabular}[c]{@{}c@{}}Diff. \&\\      Comm.\end{tabular}} \\ \cmidrule(lr){2-10}
                         & Train \#     & Valid \#    & Test \#    & Train \#       & Valid \#       & Test \#        & Train \#   & Valid \#   & Test \#    &                          &                        &                                                                                           &                                                                                \\ \midrule
Tufano                   & --            & --           & --          & $\sim$134k     & $\sim$17k     & $\sim$17k     & $\sim$134k & $\sim$17k & $\sim$17k & 1                        & Func.                  & \ding{56}                                                                                         & \ding{56}                                                                              \\
Crer                     & $\sim$226k   & $\sim$31k   & $\sim$31k  & $\sim$118k     & $\sim$10k      & $\sim$10k      & $\sim$150k & $\sim$13k  & $\sim$13k  & 9                        & Line.                  & \ding{52}                                                                                         & \ding{52}                                                                              \\ \bottomrule
\end{tabular}
\end{table*}

\subsection{Evaluation Criteria}
We employ task-specific metrics to gauge the performance of our model across the code review tasks.

For review necessity prediction, we approach it as a binary classification problem where `requiring a review' is the positive class. Thus, we use precision, recall, and F1-score as evaluation metrics to quantify the model's categorization accuracy.

For the code review comment generation and code refinement tasks, which involve response generation, we use the BLEU-4 score, which measures the overlap of $n$-grams for $n$ ranging from 1 to 4. This follows the evaluation method employed in CodeReviewer \cite{li2022automating}.

We do not use the codeBLEU metric proposed by Tufano et al.\cite{tufano2022using}, due to its incompatibility with the CRer dataset. The CRer dataset's structure and language variety make it unsuitable for this metric.

For all tasks, we consider the top-1 result, aligning with our objective of automating the code review process to lighten developers' workload by focusing on the most relevant feedback.

\subsection{Baselines}
We chose our baselines based on the tasks and datasets' unique needs. Table \ref{baseline-table} displays the chosen baselines.

\begin{table*}
\caption{Summary of the Baselines.}
\centering
\begin{threeparttable}
\begin{tabular}{@{}lccc@{}}
\toprule
\textbf{Baseline (Description)} & \textbf{Tasks} & \textbf{Datasets} & \textbf{References} \\
\midrule
Transformer-s (6-layer encoder/decoder, trained from scratch) & RCG, CR & Tuf. & \cite{vaswani2017attention, tufano2022using} \\
Transformer-b (12-layer encoder/decoder, trained from scratch) & RNP, RCG, CR & CRer & \cite{vaswani2017attention, li2022automating} \\
Tufano et al. (Pre-trained Transformer-s on their datasets) & RCG, CR & CRer, Tuf. & \cite{tufano2022using} \\
CodeT5 (Pre-trained Transformer-b model for code understanding and generation) & RNP, RCG, CR & CRer & \cite{wang2021codet5, li2022automating} \\
CodeReviewer (Pre-trained with Transformer-b and specific code-review-related pre-training tasks) & RNP, RCG, CR & CRer & \cite{li2022automating} \\
CommentFinder (Retrieval-based review comment recommendation) & RCG & CRer, Tuf. & \cite{hong2022commentfinder} \\
AUGER (Re-pre-trained T5 with extra review tags input) & RCG & Tuf. & \cite{li2022auger} \\
\bottomrule
\end{tabular}
\label{baseline-table}
\begin{tablenotes}
\footnotesize
\item Task abbreviations: RNP (Review Necessity Prediction), RCG (Review Comment Generation), CR (Code Refinement).
\item Dataset abbreviations: CRer (CRer dataset), Tuf. (Tufano dataset).
\end{tablenotes}
\end{threeparttable}
\end{table*}

We omitted the studies of \cite{thongtanunam2022autotransform, pornprasitd, zhang2022coditt5} from our baseline selection, as they are designed for smaller datasets or require additional input information.

\subsection{Implementation Details}
We implemented our approach using the xturing\footnote{https://github.com/stochasticai/xturing} and Lit LLaMA\footnote{https://github.com/Lightning-AI/lit-llama} frameworks, respectively. All experiments were conducted on NVIDIA A100-SXM4-80GB GPU platforms, with the token length limit set to 2048 and a batch size of 64. We used the AdamW optimizer and trained the models for 5 epochs for review necessity prediction and 10 epochs for both code review comment generation and code refinement tasks.

For zero-init attention prefix-tuning, we used a learning rate of 0.009, weight decay of 0.02, prefix prompt length of 10, and a prefix layer of 30. In Low-rank Adaptation (LoRA), we set the learning rate to 0.0003, weight decay to 0.01, LoRA rank to 16, and the LoRA scaling factor to 16. Ablation experiment settings may vary based on each experiment's requirements. The LoRA rank and prefix-tuning setting are based on empirical experience\cite{Tloen, zhang2023llama}. More details on the specific hyper-parameters are available in our materials.

Baseline implementation was tailored for each situation. For the CRer dataset results, we used the findings reported in the CodeReviewer paper\cite{li2022automating}, as we made no modifications to the dataset. We reproduced the CommentFinder\cite{hong2022commentfinder} results on both datasets using their publicly available code. All other baseline results were derived from their provided models.

\section{Evaluation}

In this section, we sequentially address each research question, presenting the results obtained from our experiments and drawing conclusions for each RQ. Our discussion begins with an assessment of LLaMA-Reviewer's performance, followed by an exploration of input representation's influence and the initial stage of instruction tuning. We conclude with an analysis of the implications arising from different parameter-efficient fine-tuning (PEFT) methods.

\subsection{RQ1: Evaluating the Performance of LLaMA-Reviewer}
This subsection assesses LLaMA-Reviewer's performance across each task, explains the observed results, and summarises the conclusions as key findings.

\subsubsection{(RQ1.1) Review Necessity Prediction Performance}

In our review necessity prediction experiments, we focused exclusively on the CRer dataset, as it is the only dataset that provides the necessary data for this task. The results are presented in Table \ref{Review Necessity Prediction Performance}, with the last row displaying the outcomes for LLaMA-Reviewer. The preceding rows show results derived from \cite{li2022automating}. We consider the class requiring a review as positive. Importantly, the results for LLaMA-Reviewer with prefix-tuning are not included in this task due to its rigid training structure, which is not conducive to performing classification tasks.

\begin{table*}
\caption{Results of Review Necessity Prediction on CRer Dataset.}
\label{Review Necessity Prediction Performance}
\footnotesize
\centering
\begin{tabular}{@{}l|rrrr|rrr@{}}
\toprule
\multicolumn{1}{c|}{Model}     & Layers      & \begin{tabular}[c]{@{}r@{}}Model \\      Params\end{tabular} & \begin{tabular}[c]{@{}r@{}}Trainable \\      Params\end{tabular} & \begin{tabular}[c]{@{}r@{}}Storage\\       Space\end{tabular} & Prec.      & Recall         & F1             \\ \midrule
Transformer-b               & 24          & $\sim$220M                                                   & $\sim$220M                                                       & 850M                                                          & 74.50          & 46.07          & 56.93          \\
Tufano et al.                  & 12           & $\sim$60M                                                    & $\sim$60M                                                        & 231M                                                          & 70.82          & 57.20          & 63.29          \\
CodeT5                         & 24          & $\sim$220M                                                   & $\sim$220M                                                       & 850M                                                          & 70.36          & 58.96          & 64.16          \\
CodeReviewer                   & 24          & $\sim$220M                                                   & $\sim$220M                                                       & 850M                                                          & 78.60          & 65.63          & 71.53          \\ \midrule 
\textbf{LLaMA-Reviewer (LoRA)} & \textbf{32} & \textbf{$\sim$6.7B}                                          & \textbf{$\sim$8.4M}                                              & \textbf{16M}                                                  & \textbf{60.99} & \textbf{83.50} & \textbf{70.49} \\ \bottomrule
\end{tabular}
\end{table*}

LLaMA-Reviewer achieves superior recall with a comparable F1 score, as the results show, indicating that it can identify a larger number of problematic code snippets potentially prompting discussion in the subsequent code review process. This capability is crucial for reviewers as the primary objective of code review is to uncover as many potential issues as possible. In this scenario, LLaMA-Reviewer can reduce the number of code snippets that need review after filtering, without overlooking a significant number of problematic snippets.

We obtained these results through threshold adjustment. Furthermore, with a threshold of 0.5, identical to the original generation setting, LLaMA-Reviewer achieves a precision of 88.61\%, exceeding all baselines. This performance is also meaningful in real-world scenarios, where problematic code snippets are less common than normal ones, indicating that false positives can place an additional burden on reviewers.

\subsubsection{(RQ1.2) Review Comment Generation Performance}

We evaluated the code review comment generation task using both the Tufano and CRer datasets. Table \ref{Review Comment Generation Performance} illustrates the results, with the symbol ``$-$" signifying missing values. It is noteworthy that CommentFinder\cite{hong2022commentfinder} does not include parameter numbers as it does not utilize a deep learning method. We have opted not to report certain baseline results due to incongruity in granularity between the proposed methods and the dataset, which makes the results meaningless.

\begin{table}
\begin{threeparttable}
\caption{Results of Review Comment Generation.}
\label{Review Comment Generation Performance}
\footnotesize
\centering
\begin{tabular}{@{}lrrrrrr@{}}
\toprule
\multicolumn{1}{c}{\multirow{2}{*}{Model}} & \multirow{2}{*}{L.} & \multirow{2}{*}{\begin{tabular}[c]{@{}r@{}}Model \\      Params\end{tabular}} & \multirow{2}{*}{\begin{tabular}[c]{@{}r@{}}Trainable \\      Params\end{tabular}} & \multirow{2}{*}{\begin{tabular}[c]{@{}r@{}}Storage\\       Space\end{tabular}} & \multicolumn{2}{c}{BLEU-4}    \\ \cmidrule(l){6-7} 
\multicolumn{1}{c}{}                       &                         &                                                                               &                                                                                   &                                                                                & Crer.  & Tuf.  \\ \midrule
Transformer-s                          & 12                       & $\sim$60M                                                                     & $\sim$60M                                                                         & 231M                                                                           & --             & 6.94\textsuperscript{*}          \\
Transformer-b                         & 24                      & $\sim$220M                                                                    & $\sim$220M                                                                        & 850M                                                                           & 4.76          & --             \\
Tufano et al.                              & 12                       & $\sim$60M                                                                     & $\sim$60M                                                                         & 231M                                                                           & 4.39          & 7.39\textsuperscript{*}          \\
CodeT5                                     & 24                      & $\sim$220M                                                                    & $\sim$220M                                                                        & 850M                                                                           & 4.83          & --             \\
CodeReviewer                               & 24                      & $\sim$220M                                                                    & $\sim$220M                                                                        & 850M                                                                           & 5.32          & --             \\
CommentFinder                          & --                       & --                                                                             & --                                                                                 & $\sim$100M                                                                     & 3.82\textsuperscript{*}          & 4.19\textsuperscript{*}          \\
AUGER                                      & 24                      & $\sim$220M                                                                    & $\sim$220M                                                                        & 850M                                                                           & --             & 3.03\textsuperscript{*}          \\ \midrule
\textbf{Ours (Prefix)}    & \textbf{32}             & \textbf{$\sim$6.7B}                                                           & \textbf{$\sim$1.2M}                                                               & \textbf{2.4M}                                                                  & \textbf{5.16} & \textbf{4.66} \\
\textbf{Ours (LoRA)}             & \textbf{32}             & \textbf{$\sim$6.7B}                                                           & \textbf{$\sim$8.4M}                                                               & \textbf{16M}                                                                   & \textbf{5.70}  & \textbf{5.04} \\ \bottomrule
\end{tabular}
\begin{tablenotes}
\footnotesize
\item [*] Denotes results achieved by their given code or model.
\end{tablenotes}
\end{threeparttable}
\end{table}

The results indicate that LLaMA-Reviewer surpasses all baselines on the CRer dataset, especially when employing Low-Rank Adaptation (LoRA) for fine-tuning. This superior performance highlights the potential of large language models (LLMs). Even though LLaMA wasn't specifically pre-trained for code review tasks like CodeReviewer\cite{li2022automating}, its superior performance with a limited amount of tuning outstrips that of smaller models. The results on the Tufano dataset are relatively less ideal, which we will further discuss in RQ2.1.

A plausible explanation for this enhanced performance is the congruence between the task of natural language generation and LLaMA's pre-training corpus. Furthermore, the impressive performance on the CRer dataset could be attributed to the use of code differences and the raw code format, mirroring the conditions of the pre-training stage. Given the complexity of the code review comment generation task in comparison to the other tasks \cite{li2022automating}, the larger model size of LLaMA provides a distinct advantage.

\subsubsection{(RQ1.3) Code Refinement Performance}

We evaluated the code refinement task on both the Tufano and CRer datasets. The results are shown in Table \ref{Code Refinement Performance}, where the symbol ``$-$" signifies missing values.

\begin{table}
\begin{threeparttable}
\caption{Results of Code Refinement.}
\label{Code Refinement Performance}
\footnotesize
\centering
\begin{tabular}{@{}lrrrrrr@{}}
\toprule
\multicolumn{1}{c}{\multirow{2}{*}{Model}} & \multirow{2}{*}{L.} & \multirow{2}{*}{\begin{tabular}[c]{@{}r@{}}Model \\      Params\end{tabular}} & \multirow{2}{*}{\begin{tabular}[c]{@{}r@{}}Trainable \\      Params\end{tabular}} & \multirow{2}{*}{\begin{tabular}[c]{@{}r@{}}Storage\\       Space\end{tabular}} & \multicolumn{2}{c}{BLEU-4}      \\ \cmidrule(l){6-7} 
\multicolumn{1}{c}{}                       &                     &                                                                               &                                                                                   &                                                                                & Crer.           & Tuf.           \\ \midrule
Transformer-s                          & 12                   & $\sim$60M                                                                     & $\sim$60M                                                                         & 231M                                                                           & --              & 77.54\textsuperscript{*}          \\
Tufano et al.                              & 12                   & $\sim$60M                                                                     & $\sim$60M                                                                         & 231M                                                                           & 77.03          & 78.33\textsuperscript{*}          \\
CodeT5                                     & 24                  & $\sim$220M                                                                    & $\sim$220M                                                                        & 850M                                                                           & 80.82          & --              \\
CodeReviewer                               & 24                  & $\sim$220M                                                                    & $\sim$220M                                                                        & 850M                                                                           & 82.61          & --              \\ \midrule
\textbf{Ours (Prefix)}                     & \textbf{32}         & \textbf{$\sim$6.7B}                                                           & \textbf{$\sim$1.2M}                                                               & \textbf{2.4M}                                                                  & \textbf{76.71} & \textbf{77.04} \\
\textbf{Ours (LoRA)}                       & \textbf{32}         & \textbf{$\sim$6.7B}                                                           & \textbf{$\sim$8.4M}                                                               & \textbf{16M}                                                                   & \textbf{82.27} & \textbf{78.23} \\ \bottomrule
\end{tabular}
\begin{tablenotes}
\footnotesize
\item [*] Denotes results achieved by their given code or model.
\end{tablenotes}
\end{threeparttable}
\end{table}

On both datasets, despite not outperforming all models, LLaMA-Reviewer competes closely with CodeReviewer\cite{li2022automating} or Tufano et al.\cite{tufano2022using}, the models pre-trained specifically for code review and the corresponding data format and exceeds the performance of the other baselines. Considering that we used the smallest version of LLaMA and limited tuning epochs, this outcome suggests potential improvements.

The LLaMA-Reviewer's advantage over most baselines primarily arises from its large model size and the nature of the pre-training data. The gap between LLaMA-Reviewer and CodeReviewer\cite{li2022automating} or Tufano et al.\cite{tufano2022using} is due to the differences between their target tasks and LLaMA's pre-training tasks, as well as the input formats. However, task-specific pre-training from scratch, as with CodeReviewer\cite{li2022automating}, is resource-intensive, creating a barrier for enhancement through model size expansion. Instead, integrating domain knowledge into a single pre-trained model and applying parameter-efficient fine-tuning methods could be more cost-effective.

Interestingly, the relative simplicity of the code refinement task compared to the review comment generation task may have paradoxically lowered the LLaMA-Reviewer's BLEU score. This is because the model, trained to generate diverse predictions mimicking human behavior, might produce more diverse, yet valid, refinements which diverge from the singular ground truth, thereby decreasing textual similarity.

\begin{tcolorbox}
\vspace{-0.15cm}
\textbf{Answer to RQ1}: LLaMA-Reviewer relatively more excels in generating review comments (NL) and identifies more problems in necessity prediction while maintaining competitive performance in code refinement.
\vspace{-0.15cm}
\end{tcolorbox}

\subsection{RQ2: The Influence of Input Representation}
In this subsection, we investigate the impact of input representation using the results from RQ1 and additional ablation experiments. We address each sub-question in turn before drawing an overall conclusion.

\subsubsection{(RQ2.1) Consequences of Code Formatting}
To evaluate the effect of code formatting, we examine the results of code comment generation and code refinement tasks using both the CRer and Tufano datasets.

The results presented in Section V-A show superior relative performance on the CRer dataset compared to the Tufano dataset. Despite the differences in data distribution, the primary distinction between these two datasets lies in their code formatting. The code in the CRer dataset is more rudimentary and akin to the format used during LLaMA's pre-training, while the code in the Tufano dataset has undergone sophisticated processing. These results suggest that a code representation similar to the one used during pre-training allows the model to better leverage its understanding of code structures and semantics.

\subsubsection{(RQ2.2) Role of Language Label}
We conducted experiments to evaluate the impact of language labels solely on the CRer dataset, as it includes multiple programming languages. Language labels, determined based on the programming language of the code, were integrated either into the instruction or the input, as shown in Figure \ref{fig: prompt}. The remaining settings, borrowed from the code refinement task with low-rank adaptation as a fine-tuning method, were kept constant.

Contrary to expectations, the results presented in Table \ref{Role of language label} reveal that adding a language label does not enhance performance without an initial phase of instruction tuning. This could be due to the model's difficulty in associating label information with the task without pre-existing domain knowledge. However, once instruction tuning is implemented, the labels indeed contribute positively to the model's performance. Through a paired bootstrap resampling test, we determined that the use of a language label improves performance over its absence, evidenced by a p-value of 0.0032.

Although language labels have demonstrated their value, we chose not to include them in other experiments in order to maintain consistency with previous research \cite{li2022automating} and to ensure a fair comparison.

\begin{table}
\caption{Role of Language Label (LoRA $r$ = 8).}
\label{Role of language label}
\footnotesize
\centering
\begin{tabular}{@{}l|ccc@{}}
\toprule
Model                                                                                                           & Lang. Label & Placement   & BLEU-4         \\ \midrule
\multirow{3}{*}{\begin{tabular}[c]{@{}l@{}}LLaMA-Reviewer   (LoRA) \\      w/o Instruction Tuning\end{tabular}} & \ding{56}           & --           & \textbf{81.87} \\
                                                                                                                & \ding{52}           & Instruction & 81.07          \\
                                                                                                                & \ding{52}           & Input       & 81.33          \\ \midrule
\multirow{2}{*}{\begin{tabular}[c]{@{}l@{}}LLaMA-Reviewer   (LoRA)\\      w/ Instruction Tuning\end{tabular}}   & \ding{56}           & --           & 81.59          \\
                                                                                                                & \ding{52}           & Instruction & \textbf{82.00}    \\ \bottomrule
\end{tabular}
\end{table}

\begin{tcolorbox}
\vspace{-0.15cm}
\textbf{Answer to RQ2}: LLaMA-Reviewer performs better when the input representation resembles that used during pre-training. Additional natural language information, such as language labels, can be better leveraged by the model through instruction tuning.
\vspace{-0.15cm}
\end{tcolorbox}

\subsection{RQ3: The Impact of Instruction Tuning}
To answer RQ3, we perform experiments with models trained with and without the preliminary stage of instruction tuning, employing both zero-init attention prefix-tuning and Low-Rank Adaptation (LoRA). We also introduce supplementary experiments with additional natural language instructions \cite{alpaca} during LoRA instruction tuning to identify the optimal instruction tuning data (as posed by RQ3.3). The experiments are conducted on the CRer dataset using LoRA, and the results from the code review tasks are documented in Table \ref{Impact of Instruction Tuning}.

\begin{table}
\caption{Impact of Instruction Tuning (LoRA $r$ = 8).}
\label{Impact of Instruction Tuning}
\footnotesize
\centering
\begin{tabular}{@{}l|ccccc@{}}
\toprule
Method                                                                                             & \multicolumn{1}{l}{I. Tuning} & Dataset & RNP (F1)     & RCG     & CR       \\ \midrule
\multirow{3}{*}{\begin{tabular}[c]{@{}l@{}}LoRA\end{tabular}}          & \ding{56}                             & --       & \textbf{70.20}          & 5.58          & \textbf{81.87} \\
                                                                                                  & \ding{52}                             & PL      & 69.34 & \textbf{5.64} & 81.59          \\
                                                                                                  & \ding{52}                             & PL + NL & 69.82          & 5.23          & 81.17          \\ \midrule
\multirow{2}{*}{\begin{tabular}[c]{@{}l@{}}Prefix- \\ tuning\end{tabular}} & \ding{56}                             & --       & -- & \textbf{5.16} & \textbf{76.71} \\
                                                                                                  & \ding{52}                             & PL      & --          & 5.02          & 76.04          \\ \bottomrule
\end{tabular}
\end{table}

\subsubsection{(RQ3.1) Consequences for Zero-init Attention Prefix-tuning}
Our results suggest that instruction tuning is not conducive to prefix tuning. This could be attributed to the structure of prefix tuning, which employs the prefix solely to control attention and keeps attention across the postfix fixed. As such, its ability to capture general domain knowledge is limited. Moreover, the zero-init prefix attention, which contributes significantly to the efficacy of prefix tuning, is undermined when instruction tuning is added.

\subsubsection{(RQ3.2) Consequences for Low-Rank Adaptation}

Unlike prefix tuning, instruction tuning in conjunction with LoRA improves performance across most tasks. For review necessity prediction, it elevates the precision of predictions from 81.56\% to 83.99\% when using the PL dataset as the sole tuning set, even though it doesn't enhance the f1-score. For review comment generation, it increases the BLEU score.\footnote{This effect is more pronounced with learning rate 5e-5. Post 10 epochs, the results with and without instruction tuning are 5.43 and 5.27, respectively.} For code refinement, although no substantial improvement is detected, we infer from Section V-B that it augments the model's ability to incorporate label information. Instruction tuning with LoRA helps the base model comprehend instructional intent, which, in turn, benefits subsequent task tuning, particularly review comment generation, given its complexity and multi-intent nature.

\subsubsection{(RQ3.3) Influence of Instruction Types}
We focus on the ``PL+NL" and ``PL" rows using LoRA, which denote the use of both Alpaca and Code Alpaca data and only Code Alpaca data, respectively. Interestingly, even though code review tasks are closely related to natural language, incorporating Alpaca data for instruction tuning diminishes performance across all tasks. This trend may be linked to the extensive diversity of natural language instructions in Alpaca. Vast instructions in the Alpaca dataset are with verbs like rewrite and classify, and they seem to be overwhelming for code review tasks due to their wide range of tasks.

\begin{tcolorbox}
\vspace{-0.15cm}
\textbf{Answer to RQ3}: Instruction tuning can potentially enhance task performance or the capacity to handle additional natural language information. However, the effectiveness is minor due to the word habit inconsistency among instruction and downstream datasets.
\vspace{-0.15cm}
\end{tcolorbox}

\subsection{RQ4: Influence of Parameter-Efficient Fine-Tuning}
To explore the impact of Parameter-Efficient Fine-Tuning (PEFT) methods, we conduct additional experiments adjusting the rank $r$ of LoRA, specifically setting the rank to 8 and 16. The investigation of hyper-parameters for prefix-tuning is ommited as it was fully analyzed in previous work\cite{zhang2023llama}. The outcomes are detailed in Table \ref{Influence of PEFT}. For comparative analysis, we also include the results of prefix-tuning.

\begin{table}
\caption{Influence of Parameter-Efficient Fine-Tuning Methods.}
\label{Influence of PEFT}
\footnotesize
\centering
\begin{tabular}{@{}l|rrr|rrr@{}}
\toprule
\begin{tabular}[c]{@{}l@{}}Tuning\\ Method\end{tabular} & $r$  & \begin{tabular}[c]{@{}r@{}}Trainable\\ Params\end{tabular} & \begin{tabular}[c]{@{}r@{}}Storage\\       Space\end{tabular} & \multicolumn{1}{c}{\begin{tabular}[c]{@{}c@{}}RNP\\ (F1)\end{tabular}} & \multicolumn{1}{c}{\begin{tabular}[c]{@{}c@{}}RCG\\ (BLEU)\end{tabular}} & \multicolumn{1}{c}{\begin{tabular}[c]{@{}c@{}}CR\\ (BLEU)\end{tabular}} \\ \midrule
Prefix                                                  & --  & $\sim$1.2M                                                 & 2.4M                                                          & --                                                                        & 5.16                                                                     & 76.71                                                                   \\
LoRA                                                    & 8  & $\sim$4.2M                                                 & 8M                                                            & 69.34                                                                    & 5.64                                                                     & 81.59                                                                   \\
LoRA                                                    & 16 & $\sim$8.4M                                                 & 16M                                                           & \textbf{70.49}                                                                    & \textbf{5.7}                                                                      & \textbf{82.27}                                                                  \\ \bottomrule
\end{tabular}
\end{table}

\subsubsection{(RQ4.1) Comparison between PEFT Methods}
The results decisively indicate that LoRA surpasses prefix-tuning across all tasks. We attribute this enhanced performance to two primary aspects. First, the prefix-tuning method implemented in our study has fewer trainable parameters than LoRA, impeding its adaptability from the base model. Second, in contrast to prefix-tuning which relies on prefixes to control the base model, LoRA approximates full-parameter tuning, an attribute crucial when the target output significantly differs from the pre-training format.

\subsubsection{(RQ4.2) Impact of LoRA Rank $r$}
Increasing the LoRA rank $r$ from 8 to 16 improves the performance of LLaMA-Reviewer. This improvement is intuitive, as a higher rank augments the number of trainable parameters, bringing the model closer to full-parameter tuning. Nevertheless, the chief aim of using PEFT methods is to limit trainable parameters and conserve computational resources. Thus, striking a balance between performance and efficiency is a vital consideration.

\subsubsection{(RQ4.3) Efficiency of PEFT Methods}
As shown in the table, PEFT methods reduce the number of trainable parameters to under 1\% while still ensuring acceptable performance. Given that PEFT methods keep the weights of the base model constant, the storage space decreases drastically from 13GB to under 20MB. These independent plug-in weights make PEFT methods an ideal fit for multi-task processes, such as the automation of code review activities.

\begin{tcolorbox}
\vspace{-0.15cm}
\textbf{Answer to RQ4}: Among PEFT methods, LoRA is more suitable for automating code review tasks. By choosing an appropriate LoRA rank, LLaMA-Reviewer can achieve competitive performance with less than 1\% of the trainable parameters and significantly reduced storage space requiements.
\vspace{-0.15cm}
\end{tcolorbox}

\section{Related Work}
\subsection{Tuning on LLaMA}

While OpenAI's proprietary ChatGPT and GPT series have driven substantial progress in the AI field, their closed-source nature has generated reservations among researchers. Addressing this concern, Meta introduced their open-source Large Language Model (LLaMA) \cite{touvron2023llama}, which has swiftly emerged as a pivotal asset in the AI landscape due to its remarkable performance capabilities.

A variety of LLaMA-based tuned models have shown exceptional performance, rivaling the ChatGPT and GPT series. Stanford's Alpaca \cite{alpaca} represents an early significant development in this area, tuning LLaMA using a dataset generated from ChatGPT. Subsequent notable works have pursued a range of objectives, including language-specific adaptations \cite{leng2023chinese-vicuna, chinese-llama-alpaca}, text quality enhancement \cite{vicuna2023, peng2023instruction}, multi-modal input accommodation \cite{zhang2023llama, zhu2023minigpt, liu2023llava}, and code-related capabilities improvement \cite{codealpaca}.

Due to the substantial computational demands of full parameter tuning, researchers have gravitated towards Parameter-Efficient Fine-Tuning (PEFT) methods for tuning LLaMA. One such example is Alpaca LoRA, which achieves performance comparable to Alpaca with only 0.13\% of the training parameters, resulting in approximately a 60 times speedup \cite{Tloen}. LLaMA-adapter, introduced by Zhang et al. \cite{zhang2023llama}, represents a further contribution, employing a zero-init attention prefix-tuning method.

Our study focuses on evaluating LLaMA's performance on code review-related tasks and employs state-of-the-art PEFT methods to enhance training efficiency.

\subsection{Automation of Code Review Activities}
Code review is an essential, albeit time-consuming, aspect of software development, sparking considerable interest in automation strategies for activities such as reviewer recommendation \cite{sulun2019suggesting, asthana2019whodo, chueshev2020expanding, rebai2020multi, mirsaeedi2020mitigating, sulun2021rstrace+, gauthier2021historical, kong2022recommending, rong2022modeling, pandya2022corms}, code quality assessment \cite{shi2019automatic, hellendoorn2021towards, hijazi2021ireview, wang2021can, gauthier2021historical, li2022automating, hijazi2022quality}, problematic code refinement \cite{tufano2021towards, tufano2022using, fu2022vulrepair, li2022automating, thongtanunam2022autotransform}, and review comment suggestion \cite{balachandran2013reducing, gupta2018intelligent, siow2020core, hong2022should, tufano2022using, li2022automating, li2022auger, hong2022commentfinder}. This paper focuses on the pipeline proposed by Li et al. \cite{li2022automating}, which comprises review necessity prediction, code review comment generation, and code refinement.

Early studies on review necessity prediction primarily examined diff hunk acceptance, with Shi et al. \cite{shi2019automatic} pioneering a CNN and LSTM-based framework, DACE, and Hellendoorn et al. \cite{hellendoorn2021towards} utilizing a Transformer to account for inter-diff hunk relations within a single pull request (PR). However, the field has since evolved, with Li et al. \cite{li2022automating} shifting focus towards identifying diff hunks requiring review and integrating this into a pipeline with a code review-specific pre-trained model.

Initial efforts in code review comments leveraged retrieval-based methods for historical comment extraction. Gupta et al. \cite{gupta2018intelligent} introduced an LSTM-based model, DeepMem, to recommend comments for new code snippets based on their relationship with code changes, while Siow et al. \cite{siow2020core} enhanced retrieval through attention-based LSTM semantic information capture. The field has since shifted towards generating review comments with the rise of deep learning. Tufano et al. \cite{tufano2022using} pioneered this approach, pre-training a model on both code and technical language, with subsequent efforts by CodeReviewer \cite{li2022automating} and AUGER \cite{li2022auger} using a code review-specific pre-trained model and review tags, respectively, for improved results. At the same time, CommentFinder \cite{hong2022commentfinder} presents an efficient retrieval-based alternative.

For code refinement, early efforts are often aligned with automatic bug-fixing techniques \cite{tufano2019empirical, gupta2017deepfix, chen2019sequencer}. Pioneering the adaptation of this task to code review, Tufano et al. \cite{tufano2019learning} focused on learning from code changes implemented in PRs. Later researchers incorporated code review comments into task input to better simulate code refinement \cite{tufano2021towards, tufano2022using}. Addressing the challenge of new tokens, AutoTransforms \cite{thongtanunam2022autotransform} employed a byte-pair encoding (BPE) approach, additionally using a diff-aware method, D-ACT \cite{pornprasitd}, to boost performance in cases closely related to single differences between the code base and initial commit. However, they did not account for the influence of code review comments and excluded data with similar input. CoditT5 \cite{zhang2022coditt5}, a pre-trained model explicitly designed for editing, used part of Tufano's dataset for validation as a downstream task. Similarly, CodeReviewer \cite{li2022automating} developed a model based on their pre-trained model, specifically tailored for the code review process.

Despite significant progress in automating code review tasks, previous research often neglects the potential of unified large language models (LLMs). As the model size and training data continue to grow, unified LLMs are improving their performance at a rapid pace and show comparable performance to those task-specific pre-trained models. For task-specific pre-trained models, building from scratch is resource-intensive and time-consuming. Thus, in this study, we leverage LLaMA, a mainstream unified large language model, to investigate the evolving potential of LLMs and assess their suitability for tasks that encompass both programming and natural language, such as code review tasks.

\section{Threats to Validity}

\subsection{Construct Validity}
Our evaluation relies primarily on a variant of the BLEU-4 metric. Although widely used in prior research \cite{tufano2021towards,tufano2022using,li2022automating,hong2022commentfinder,zhang2022coditt5,thongtanunam2022autotransform}, it is not universally recognized as the definitive metric for assessing code review comments and refined code snippets. Other metrics such as rouge were not considered because several of the baseline models neither provide the direct result nor the fine-tuned model and generated predictions. Our tests on the baselines, including AUGER \cite{li2022auger} and CommentFinder \cite{hong2022commentfinder}, relied on the code or the model furnished by the original paper, which may have deviated from optimal results due to varying data formats and distributions.

\subsection{Internal Validity}

Our assessment of parameter-efficient fine-tuning (PEFT) was limited to two prominent PEFT methods and did not include a full-parameter fine-tuning comparison. This was due to the significant computational resources required for full-parameter fine-tuning, which necessitates 8 A100-SXM4-80G over half a month. Likewise, our ablation experiments covered a limited set of settings due to resource constraints. Although these limitations may have resulted in alternative findings, they do not fundamentally undermine our primary objective: to evaluate the potential of unified large language models.

Another potential internal validity threat is that we constrained our training to the smallest size of LLaMA and a finite number of epochs. Given that LLaMA's capabilities and data results continue to improve with an increase in model size and tuning duration, our study may understate the actual latent capacity of large language models.

\subsection{External Validity}

Our findings may not generalize beyond the context of the two datasets \cite{li2022automating,tufano2022using} used in this study, which is exclusively derived from open-source projects. As such, these findings may not fully apply to industrial and other contexts. Additionally, each dataset retains only a single comment per code change, potentially introducing bias during filtering.

\section{Conclusion and Future Work}

In this paper, we introduced LLaMA-Reviewer, a framework for automating the code review process using large language models (LLMs) and parameter-efficient fine-tuning (PEFT) techniques. We demonstrated that, despite using the smallest version of LLaMA with only 6.7B parameters and less than 1\% of trainable parameters over a limited number of tuning epochs, LLaMA-Reviewer can match the performance of state-of-the-art code-review-focused models. Additionally, by applying plug-in type models, we significantly reduce storage space requirements.

Our findings also suggest that aligning the input representation with the format used during pre-training could better leverage the capabilities of LLMs. Additionally, an initial stage of instruction tuning can improve task performance and increase the model's ability to process additional natural language information. Our results also showed that low-rank adaptation with an appropriate rank is preferable for tasks with specific input and output formats.

Looking ahead, we aim to broaden our exploration of large language models, considering models of various sizes and types, and further investigate PEFT methodologies. We are also interested in examining more closely the relationship between the token length of prompt templates, code snippets, comments, and the sequence block of the pre-trained models.

\section*{Acknowledgment}
This work was supported by Chinese Academy of Sciences-Dongguan Science and Technology Service Network Plan (No.202016002000032), and Alliance of International Science Organizations (No. ANSO-CR-KP-2022-03).

\bibliographystyle{IEEEtran}
\bibliography{base}


\end{document}